\documentclass{article}
\usepackage{amsmath,amsthm}
\usepackage{amssymb,latexsym}
\usepackage[mathscr]{eucal}
\usepackage{setspace}
\usepackage{graphics}
\usepackage{array}

\newcommand{\pa}{\partial}
\newcommand{\pr}{\prime}
\newcommand{\al}{\alpha}

\newcommand{\eps}{\epsilon}
\newcommand{\lp}{\left(}
\newcommand{\rp}{\right)}
\newcommand{\lb}{\left[}
\newcommand{\rb}{\right]}

\newcommand{\la}{\left<}
\newcommand{\ra}{\right>}
\newcommand{\be}{\begin{equation}}
\newcommand{\ee}{\end{equation}}
\newcommand{\bs}{\boldsymbol}

\newcommand{\ihat}{\bf\hat{i}}

\begin{document}

\title{\textbf{Superfluid Turbulence in the Kelvin Wave Cascade Regime}}         
\author{Bhimsen K. Shivamoggi\footnote{Permanent Address: University of Central Florida, Orlando, FL 32816-1364, USA}\\
International Centre for Theoretical Sciences (ICTS-TIFR)\\
TIFR Centre Building, IISc Campus\\
Bengaluru 560012, India
}        
\date{}          
\maketitle

\large{\bf Abstract}

Theoretical considerations are made of superfluid turbulence in the Kelvin wave cascade regime at low temperatures $(T < 1K)$ and length scales of the order or smaller than the intervortical distance. The energy spectrum is shown to be in accord with the Kolmogorov scaling. The vortex line decay equation is shown to have an underlying Hamiltonian framework. Effects of spatial intermittency (exhibited in laboratory experiments) on superfluid turbulence are incorporated via the fractal nature of the vortex lines, for length scales of the order or smaller than the intervortical distance. The spatial intermittency effects are shown to enhance the vortex line density $L$, for a given value of intervortex spacing $\ell$, and to provide for a mechanism commensurate with the enhanced depolarization of vortex lines. The spatial intermittency is found to steepen the energy spectrum in qualitative agreement with laboratory experiments and to enhance vortex line decay.

\pagebreak

\noindent\Large\textbf{1. Introduction}\\

\large Superfluids execute flows which are in principle quite distinct from their classical counterparts due to constraints imposed by long-range quantum order. Consquently, all vorticity is confined to topological defects like lines and the circulation around each defect is quantized (Onsager \cite{Ons}). Quantized vortex lines are the only excited degrees of freedom in a superfluid\footnote{On length scales larger than the intervortical distance, the vortex lines can be organized into polarized bundles to mimic classical flow patterns (ex: a uniform array of parallel vortex lines would mimic the uniform rotation of the superfluid component with the containing vessel).}. Thanks to the circulation quantization constraint on the vortex lines in superfluid $^4$He, the only possible turbulent motion in the latter system, as Feynman \cite{Fey} brilliantly envisioned, is a disordered motion of tangled vortex lines. This dynamic tangle of vortex lines and hence the intensity of the resulting superfluid turbulence is characterized by the vortex line density $L$ (which is the total length per unit volume and is the quantity usually measured in laboratory experiments and numerical simulations on liquid He) and is sustained by the mutual friction\footnote{The mutual friction is known (Schwarz \cite{Sch}, Shivamoggi \cite {Shi}, \cite{Shi2}) to play the dual roles of driving force and drag force and hence to produce both growth and decay of the vortex line length.}(Feynman \cite{Fey}, Hall and Vinen \cite{Hal}, \cite{Hal2}) due to the relative motion between the superfluid and normal fluid components\footnote{The natural motion of the vortex lines generates high-curvature regions which move at a high-speed relative to the normal fluid and scatter a lot of thermal excitations constituting the normal fluid and hence experience a large friction force that tends to reduce the curvature (Vinen and Niemela \cite{Vin}) and makes the vortex tangle self-sustained (Schwarz \cite{Sch}). Numerical simulations (Tsubota et al. \cite{Tsu}) confirmed the smoothing of the vortex lines caused by the mutual friction.}. The numerical simulations of Schwarz \cite{Sch} reduce superfluid dynamics to the tracking of a set of vortex lines which evolve as per the vortex-induced flow velocity given by the local induction approximation (LIA) (Da Rios \cite{DaR}, Arms and Hama \cite{Arm}) of the Biot-Savart's law as well as friction force\footnote{The theoretical formulation (Shivamoggi \cite{Shi}, \cite{Shi2}) of the self-advection of the vortex filament in a superfluid within the LIA framework with the Hall-Vinen-Bekarevich-Khalatnikov (\cite{Hal}, \cite{Hal2}, \cite{Bek}) phenomenological model for the mutual friction confirms the dual role of the latter as a driving force and a drag force (Schwarz \cite{Sch}). The mutual friction also provides for a mechanism to stretch the vortex lines (which are inextensional in the LIA).}. The non-local terms, neglected in LIA, become important as an element of vortex filament closely approaches another element, as in the vortex reconnection  process - this process has been observed experimentally (Bewley et al. \cite{Bew}) and is believed to be essential in randomizing the geometry of  the vortex tangle and sustaining it in a homogeneous and isotropic state.

Experimental evidence and numerical simulations (Kobayashi and Tsubota \cite{Kob}) strongly suggest (Vinen and Niemela \cite{Vin}, Skrbek and Sreenivasan \cite{Skr}, Nemirovskii \cite{Nem}) that, on large scales, quantum effects that characterize superfluid behavior become unimportant and superfluid flows exhibit coarse-grained quasi-classical behavior\footnote{Superfluid flows typically show quasi-classical behavior if the two fluid components remain locked (as at large length scales) or if the temperature is so low that the normal fluid is essentially absent (Vinen \cite{Vin2}) and decoupled as well with the superfluid component.}. On length scales much larger than the average distance between the vortices (called the intervortical distance, which provides the natural quantum length scale) $\ell$,
\be\tag{1}
\ell \sim L^{-1/2}
\ee
many vortex lines participate in the dynamics and, if they are sufficiently polarized, their collective coherent behavior materializes and the superfluid is found to support quasi-classical turbulence\footnote{If the vortex lines are totally depolarized (i.e., they are arranged in a truly random manner), then there can be no large-scale motion. However, a partial polarization with some degree of alignment can lead to large-scale motions though with vorticity not exceeding $\kappa/\ell^2$, $\kappa \equiv h/m_4$ being the quantum of circulation.} in which the energy cascades via local nonlinear interactions in the spectral space towards smaller scales until it is dissipated. The dissipation for length scales of the order or less than $\ell$ is provided by the scattering of the thermal excitations in the superfluid by the vortices. Though the viscosity of a superfluid is zero, for length scales larger than $\ell$, the superfluid is nearly locked due to mutual friction\footnote{The mutual friction between the two fluid components (mediated by a vortex tangle frozen in the superfluid component) mimics to some extent the viscous action in the cascade at $T > 1K$ and leads to strong coupling of the two fluid components which then behave effectively like a single fluid.} with the normal fluid and hence inherits an effective kinematic viscosity $\nu^\pr$ from the latter even down to the lowest temperature\footnote{The dissipative processes underlying $\nu^\pr$ include therefore not only the normal-fluid viscosity but also the mutual friction.}. This scenario has been confirmed by laboratory experiments (Maurer and Tabeling \cite{Mau}, Roche et al. \cite{Roc}, Salort et al. \cite{Sal}, \cite{Sal2}, Bradley et al. \cite{Bra}) which demonstrated that there was no observable difference in the behavior between the normal and superfluid phases and confirmed a quasi-classical turbulence with the $k^{-5/3}$ scaling for the energy spectrum\footnote{In laboratory experiments and numerical simulations, the vortex line density $L$ must be large enough to provide an inertial range sufficiently large to support the energy spectrum.} associated with large-scale polarized bundles of vortex lines; the latter was found to be independent of temperature down to $T = 1.4K$.

On the other hand, laboratory experiments on decaying grid-flow turbulence above $T = 1.1K$ (Stalp et al. \cite{Sta}) suggested and Babuin et al. \cite{Babu} confirmed that the rate of energy dissipation per unit mass, on length scales of the order of $\ell$, caused by the vortex lines is given on phenomenological grounds by
\be\tag{2}
\eps \sim \nu^\pr \kappa^2 L^2.
\ee
Vinen \cite{Vin3} pointed out that one may loosely take $\kappa^2 L^2$ as a measure of the total mean square vorticity in the superfluid component\footnote{Mean square vorticity in a superfluid is however not a very well defined quantity because vortex filaments in a superfluid are infinitely thin.} due to random vortex tangle (Vinen \cite{Vin3}),
\be\tag{3}
\la \Omega^2 \ra \sim \kappa^2 L^2.
\ee
Since (3) implies $<\Omega^2>\sim \kappa^2\ell^{-4}$, as per footnote 6, (2) underscores the presence of at least a partial vortex line polarization in this system. On the otherhand, (3) allows (2) to have the same form ({\it albeit} in a superficially similar way) as that describing viscous dissipation in classical hydrodynamic turbulence\footnote{There is some issue about interpreting the vortex line density $L$ as a measure of vorticity because the spectrum of $L$ is found in experiments (Roche et al. \cite{Roc}, Bradley et al. \cite{Bra2}) not to increase with $k$ (as it should, if the above interpretation is correct).}. (2) mimics quasi-classical behavior, on length scales of the order of $\ell$, and has also been verified qualitatively by the measurements of the vortex line density $L (t)$ in decaying superfluid turbulence (Walmsley et al. \cite{Wal}).

On length scales of the order of $\ell$, discrete vortex line effects materialize and self-advection of a vortex line controls the dynamics while dissipative processes continue to arise partly from the normal-fluid viscosity and partly from mutual friction because the velocity fields of the two fluid components are not the same. However, for $T < 1K$, the normal fluid component disappears and mutual friction becomes vanishingly small; so $\nu^\pr$ drops sharply (Walmsley et al. \cite{Wal}). However, laboratory experiments (Davis et al. \cite{Dav}) indicated the presence of other dissipative mechanisms causing the vortex line decay. Phonon radiation was ruled out as a viable dissipative mechanism for length scales of the order of $\ell$ (Vinen \cite{Vin3}, \cite{Vin4}) because it is ineffective at these length scales and is hence inadequate in accounting for the observed decay of superfluid turbulence (Vinen and Niemela \cite{Vin}). On the other hand, phonon radiation had to be rejected as a viable dissipative mechanism for low temperatures as well because of the observed (Davis et al. \cite{Dav}) temperature-independent vortex line decay rate below $T = 70 mK$ {\it along} with the strong temperature dependence of the phonon density $\lp \sim T^{-3} \rp$ in this range. Dissipation via Kelvin wave cascade has therefore been considered as a viable possibility at very low temperatures (Vinen and Niemela \cite{Vin}). This scenario may be understood by noting that the absence of the smoothing effect of the mutual friction at very low temperatures leads to sharp distortions of vortex lines like cusps and kinks\footnote{The kinkiness of the vortex lines is indeed found to increase as temperature decreases (Schwarz \cite{Sch}).} which become seats of vortex self-reconnection\footnote{Vortex reconnections allow the vortex tangle to evolve toward a lower-energy configuration via vortex line decay and energy is transferred to the normal fluid via mutual friction or to the phonon or Kelvon generation (Paoletti et al. \cite{Pao2}).} and fragmentation (Kozik and Svistunov \cite{Koz}) and generate Kelvin waves on individual vortex lines\footnote{Kelvin waves are circularly-polarized waves and are associated with helical displacements of the vortex cores in an inviscid fluid which were theoretically predicted in the 19th century (Thomson \cite{Tho}). However, an experimental confirmation had to wait until the discovery of quantized vortices in superfluid $^4$He and they were observed in a uniformly rotating superfluid $^4$He (Hall \cite{Hal3}).}, as confirmed by recent laboratory experiments (Fonda et al. \cite{Fon}). The Kelvin waves are believed to interact nonlinearly, but locally in the spectral space, to produce Kelvin waves at higher frequencies, and hence creating a Kelvin-wave cascade (Svistunov \cite{Svi})\footnote{This has been numerically confirmed by Kivotides et al. \cite{Kiv}.} which replaces the Kolmogorov type cascade operational at higher temperatures. In this cascade, energy is carried from length scales of the order of $\ell$ to smaller and smaller scales by Kelvin waves on the individual vortex lines until it is dissipated via phonon radiation (Nazarenko \cite{Naz}, Vinen \cite{Vin5})\footnote{In the numerical simulations by Vinen et al. \cite{Vin6}, Kelvin waves were excited on a vortex line in superfluid $^4$He at very low temperature (so the Kelvin waves suffer negligible damping due to mutual friction with normal fluid) by continuously driving the system at a small wavenumber. The excited Kelvin wave generated higher frequency modes via nonlinear coupling which then dissipated via phonon radiation. A steady state was established (as predicted by Svistunov \cite{Svi}, by analogy with the Kolmogorov type cascade) showing a Kelvin wave cascade which was insensitive to the details of the drive.}\textsuperscript{,}\footnote{There is, however, as yet no tangible laboratory experimental evidence for the existence of a Kelvin-wave cascade (Eltsov et al. \cite{Elt}, Vinen \cite{Vin7}); direct measurements of phonon radiation from individual vortex lines need to be done (Vinen \cite{Vin8}).}.

On the other hand, for $T < T_\lambda$, the normal fluid component essentially vanishes, while for length scales less than $\ell$, the Kelvin waves govern the dynamics so at these low temperatures and small length scales superfluid turbulence may be expected to be very different from classical turbulence. However, laboratory experiments on superfluid turbulence for $T < T_\lambda$ in a cryogenic helium wind tunnel (Salort et al. \cite{Sal3}) as well as numerical simulations (Araki et al. \cite{Ara}) of superfluid turbulence without normal fluid using the vortex filament model have indicated otherwise. This has posed a major issue for fitting the experimental facts into a proper theoretical scheme on this problem (Nemirovskii \cite{Nem}), a general operability of a Kolmogorov type turbulence in any nonlinearly interacting dissipative system of many scales notwithstanding (Procaccia and Sreenivasan \cite{Pro})\footnote{Numerical simulations on turbulence in a Bose-Einstein condensate described by the Gross-Pitaevskii equation (Yepez et al. \cite{Yep}) gave a non-Kolmogorov type scaling. However, there is some controversy about connecting this result with Kelvin wave turbulence (L'vov and Nazarenko \cite{L'v}, Krstulovic and Brachet \cite{Krs}).}.

Further, laboratory experiments (Maurer and Tabeling \cite{Mau}\footnote{The laboratory experiments \cite{Mau} pertain to the quasi-classical turbulence regime prevalent for length scales longer than $\ell$.}, Salort et al. \cite{Sal3}) gave evidence of inertial range spatial intermittency in superfluid turbulence - velocity gradient probability density function (PDF) shows non-Gaussianity while structure function exponents show deviation from the Kolmogorov scaling. The laboratory experiments of Paoletti et al. \cite{Pao} showed that even the velocity field exhibited non-Gaussian statistics\footnote{The laboratory experiments of Salort et al. \cite{Sal3} indicated otherwise and the discrepancy is not resolved.}. On the other hand, thanks to excessive crinkling operational at length scales smaller than $\ell$, the vortex lines like self-avoiding lines of polymers (de Gennes \cite{deG}), are not smooth in this range (Tsubota et al. \cite{Tsu2}, Vinen \cite{Vin2}). One may therefore follow Mandelbrot \cite{Man} and argue that the spatial intermittency effects in superfluid turbulence are related to the fractal nature of the vortex lines, for length scales of the order or smaller than $\ell$\footnote{The fractalization of vortex lines in superfluid turbulence, for length scales of the order of $\ell$, was originally proposed by Svistunov \cite{Svi}. It may be mentioned that the fractalization of vortex lines for length scales much larger than $\ell$ was confirmed via numerical simulations by Baggaley and Barenghi \cite{Bag}, while Nemirovskii et al. \cite{NTA}, following Passot et al. \cite{Pass} for classical turbulence, tried to use a fractalized vortex tangle model which did not give the Kolmogorov scaling.}, and use the fractal properties of the vortex lines to determine the energy spectrum.

The purpose of this paper is to do theoretical considerations to shed light on the classical like behavior, contrary to common expectation, of superfluid turbulence in the Kelvin-wave cascade regime at low temperatures and length scales of the order or smaller than $\ell$. Spatial intermittency effects are incorporated into the theoretical formulations via the fractal nature of the vortex lines. A Hamiltonian framework underlying the vortex line decay process is exhibited.

\vspace{.3in}

\noindent\Large\textbf{2. Kelvin Wave Cascade}\\

\large The vortex tangle in superfluid turbulence is believed to undergo repeated reconnection processes generating Kelvin waves continually in the process. The Kelvin waves would then interact nonlinearly but locally in the spectral space to produce Kelvin waves at higher frequencies and hence creating a Kelvin-wave cascade. One may then consider energy to be fed into the Kelvin waves near a length scale of the order of $\ell$ which would then cascade smoothly through nonlinear processes to smaller length scales (Walmsley et al. \cite{Wal2}) until it is dissipated via phonon radiation (Vinen \cite{Vin5}). One may then consider for the energy (or smoothed vortex line density (Svistunov \cite{Svi})) cascade in superfluid turbulence an inertial range of quasi-Kolmogorov type (Kozik and Svistunov \cite{Koz}) which is assumed to be in a state of statistical quasi-equilibrium\footnote{Though such a stipulation, because of the absence of vortex stretching in the Kelvin wave cascade regime, may superficially appear not to be on strong grounds, it is pertinent to note that a similar stipulation successfully applied to the two-dimensional fluid turbulence problem where there is no vortex stretching either.}.

If the wavelength of the Kelvin waves is large compared with the vortex core radius $a$, the leading contribution to the Kelvin wave dynamics is given by the LIA (Kozik and Svistunov \cite{KozSvi}). The LIA was used as the basis also for stochastic Kelvin wave dynamics (Sonin \cite{Son}), and the latter process pertaining to a single vortex line has been argued (Kozik and Svistunov \cite{KozSvi}) to be adequate for the determination of the energy spectrum. Using the Kelvin wave dispersion relation (Donnelly \cite{Don}), (in usual notation),
\be\tag{4}
\omega = \pm \kappa ~\ell n \lp <R>/a \rp k^2
\ee  

\noindent $<R>$ being the average radius of curvature of the vortex line\footnote{Thanks to the weak logarithmic dependence, the actual value of $<R>$ is not expected to have a significant influence on (4).}. Taking\footnote{The rationale behind a Kelvin wave cascade scenario governed self-consistently only by the single scale $\ell$ has also been emphasized recently by Sonin \cite{Son}. The average radius of curvature $<R>$ is precluded from introducing a new length scale into the problem by assuming it to be comparable in magnitude to the intervortical distance $\ell$ (Schwarz \cite{Sch}) - this situation corresponds to the self-induced velocity of the vortex becoming comparable to the velocity induced by the neighboring vortex. Indeed, in a laboratory experiment, a single passage of a superfluid through a grid may be expected to generate turbulence only if it exists on length scales $\sim O\lp\ell\rp$ (Vinen \cite{Vin8}).} $k \sim \ell^{-1}$, the characteristic velocity on a length scale $\ell$ is (Vinen \cite{Vin9}),
\be\tag{5}
v (\ell) \sim \frac{\kappa}{\ell}
\ee
and hence the energy per unit mass at length scale $\ell$ is
\be\tag{6a}
E (\ell) \sim \frac{\kappa^2}{\ell^2}.
\ee

Noting that the characteristic time at length scale $\ell$ is (Smith et al. \cite{Smi}, Svistunov \cite{Svi}),
\be\tag{7}
t (\ell) \sim \frac{\ell^2}{\kappa}
\ee
the rate of energy transfer per unit mass at length scale $\ell$ is
\be\tag{8a}
\eps (\ell) \sim \frac{E (\ell)}{t (\ell)} \sim \frac{\kappa^3}{\ell^4}.
\ee

In the inertial range, we assume a quasi-stationary process in which the energy transfer rate (or a smoothed vortex line density flux (Svistunov \cite{Svi})) is nearly constant\footnote{Laboratory experiments (Walmsley et al. \cite{Wal}) on superfluid turbulence produced by an impulsive spin-down process showed a steady state inertial cascade with a constant energy flux down the range of length scales.},
\be\tag{9}
\eps (\ell) \sim const = \eps.
\ee
Using (8a), (9) leads to
\be\tag{10}
\ell \sim \frac{\kappa^{3/4}}{\eps^{1/4}}.
\ee
It is of interest to note that (10) was found empirically via direct numerical simulations of superfluid turbulence (Salort et al. \cite{Sal2}). (10) implies that the intervortical distance $\ell$ mimics the Kolmogorov microscale with the viscosity $\nu$ replaced by the quantum circulation $\kappa$\footnote{The dissipation in the Kelvin-wave cascade was shown to be characterizable via an effective kinematic viscosity (Vinen \cite{WFV3}), operating on the scale $\ell$, (see also Skrbek and Sreenivasan \cite{Skr}).} (see also (16) below).

Using (10), (6a) becomes
\be\tag{11}
E (\ell) \sim \eps^{2/3} \ell^{2/3}
\ee
which leads to the Kolmogorov energy spectrum,
\be\tag{12}
E (k) \sim \eps^{2/3} k^{-5/3}
\ee
observed in laboratory experiments (Salort et al. \cite{Sal3}) and numerical simulations (Araki et al. \cite{Ara})\footnote{The energy spectrum (12) is also in agreement with weak Kelvin wave superfluid turbulence theory given by L'vov and Nazarenko \cite{L'v2}, Bou\'{e} et al. \cite{Bou}.}. If the turbulent state in superfluid $^4He$ is considered to be the disordered motion of tangled vortex lines (Feynman \cite{Fey}), (12) appears to become inevitable since  the generation of vorticity in superfluid $^4He$ signifies local destruction of superfluidity aspects (Kaganov and Lifshitz \cite{Kag}). On the other hand, using (5), the vorticity at length scale $\ell$ is given by
\be\tag{13}
\Omega (\ell) \sim \frac{v}{\ell} \sim \frac{\kappa}{\ell^2}
\ee
so the mean square vorticity is given by
\be\tag{14}
\la \Omega^2 \ra \sim \frac{\kappa^2}{\ell^4}
\ee
which shows that the mean square vorticity diverges as $\ell \Rightarrow 0$ (as also noted by Vinen \cite{Vin3}).

Using (1), (14) may be rewritten as
\be\tag{15}
\la \Omega^2 \ra \sim \kappa^2 L^2
\ee
in agreement with Vinen's \cite{Vin3} suggestion (3). (15) also underscores the presence of a partial vortex line polarization in the Kelvin wave cascade mechanism.

Further, (6a) and (8a) may also be rewritten as
\be\tag{6b}
E (L) \sim \kappa^2 L
\ee
\be\tag{8b}
\eps (L) \sim \kappa^3 L^2.
\ee

(6b) implies that the vortex line density $L$ may be roughly used as a measure of the total kinetic energy per unit volume for superfluid turbulence (Svistunov \cite{Svi}), which is plausible because the superfluid kinetic energy is associated with vortices, which are inextensional in the LIA model. On the other hand, comparison of (8b) with (2) leads to an effective kinematic viscosity\footnote{The characterization of the energy dissipation via an effective kinematic viscosity roughly equal to the quantum of circulation, even at low temperatures where the energy dissipation mechanism is different, was conjectured by Vinen \cite{Vin11}. Vinen \cite{Vin11} conjectured further that the energy dissipation is probably rather insensitive to the details of the dissipation mechanism for length scales less than $\ell$.},
\be\tag{16}
\nu^\pr \sim \kappa.
\ee
(16) was confirmed by laboratory experiments (Walmsley and Golov \cite{Wal3}) and numerical simulations (Tsubota et al. \cite{Tsu2}) at low temperatures $(T < 0.5K)$ where the usual dissipative processes (via coupling to the normal fluid) are not operational and underscores the concept of a superfluid Reynolds number (Volovik \cite{Vol}) where the circulation quantum plays the role of the kinematic viscosity.

\vspace{.3in}

\noindent\Large\textbf{3. Vortex Line Decay Equation}\\

\large Unlike the case with an ordinary fluid, the circulation quantization constraint in a superfluid makes it impossible for a vortex line to relax by gradually slowing down when subjected to dissipative processes. The vortex line tends to relax instead by reducing its total length (Vinen \cite{Vin9}). The physical mechanism driving the vortex line decay at low temperatures is believed (Schwarz \cite{Sch}, Vinen and Niemela \cite{Vin}) to be vortex reconnection leading to vortex line shrinkage and fragmentation. Svistunov \cite{Svi} indeed proposed that vortex reconnection, which causes vortex line decay, constitutes the mechanism underlying the vortex line density cascading process in Kelvin wave turbulence.

Substituting (6b) and (8b) into the relation,
\be\tag{17}
\frac{d E}{d t} = -\eps
\ee
we obtain for the vortex line decay (Vinen \cite{Vin9}),
\be\tag{18}
\frac{d L}{d t} \sim - \kappa L^2.
\ee
(18) embodies Vinen's \cite{Vin10} assumption of the self-sustained state of the superfluid vortex tangle implying that the vortex line decay rate is primarily determined by the instantaneous value of the vortex line density L. (18) confirms that this assumption is valid in the stationary and homogeneous situation underlying the present development\footnote{This assumption, however, presupposes that other structure parameters of the vortex tangle evolve much faster than the vortex line decay process (Nemirovskii \cite{Nem}).}. (18) further gives,
\be\tag{19}
L (t) \sim t^{-1}.
\ee
(19) was confirmed by laboratory experiments (Walmsley and Golov \cite{Wal3}) and numerical simulations (Tsubota et al. \cite{Tsu2}) at low temperatures $(T < 0.5K)$.

On the other hand, using (1), (18) may be re-expressed as (Svistunov \cite{Svi}),
\be\tag{20}
\frac{d \ell}{d t} \sim \kappa \ell^{-1}
\ee
from which,
\be\tag{21}
\ell (t) \sim t^{1/2}
\ee
(21) was confirmed by the laboratory observations of the reconnection process of quantized vortices (Fonda et al. \cite{Fon}) and implies that the intervortical spacing increases as the vortex lines decay, as to be expected.

In view of the conservative nature of the mechanism underlying vortex line decay at low temperatures described by (18), it is pertinent to inquire if the latter can be characterized via a Hamiltonian framework, as explored in Section 4.

\vspace{.3in}

\noindent\Large\textbf{4. Hamiltonian Formulation for the Vortex Line Decay Equation}\\

\large Note that the motion of vortex lines at low temperatures, where mutual friction is vanishingly small, is given by the Biot-Savart law,
\be\tag{22}
\dot{\bf{s}} = \frac{\kappa}{4 \pi} \int \frac{\lp {\bf s}_0 - {\bf s} \rp \times d {\bf s}_0}{| {\bf s}_0 - {\bf s} |^3}
\ee
where ${\bf s}_0 = {\bf s}_0 (\xi, t)$ prescribes the vortex line, and ${\bf s}$ is a field point while ${\bf s}_0$ is a source point and a variable location on the vortex line. The motion given by equation (22) complies two constants,\\
\indent kinetic energy:
\be\tag{23}
E = \int \int \frac{d {\bf s} ~d {\bf s}_0}{| {\bf s} - {\bf s}_0 |} = \frac{1}{2} \int {\bf A} \cdot {\bs \Omega} ~d {\bf s}
\ee
\indent momentum:
\be\tag{24}
{\bf P} = \int {\bf s} \times d {\bf s} = \frac{1}{2} \int {\bf s} \times {\bs \Omega} ~d {\bf s}
\ee
on appropriately normalizing the variables. Here, {\bf A} is the Stokes' stream function. It may be noted that {\bf P} is also the Lamb {\it fluid impulse} integral (Batchelor \cite{Bat}).

If one uses the vortex line density as a measure of the total kinetic energy for a superfluid (which is totally valid in the LIA (Svistunov \cite{Svi})), the relaxation process for a vortex line in a superfluid may be viewed to have a variational character - minimizing $E$ while keeping {\bf P} fixed ({\bf P} not being sign definite) - a kind of {\it Beltramization} process (see Shivamoggi \cite{Shi3}).

In order to see the Hamilton equation perspective on (18), consider the vortex line in the form of an axisymmetric vortex ring of toroidal radius $R$ and vorticity $\Omega_\theta$, with ${\bf i}_z$ as the unit vector along the axis of the vortex ring. (23) and (24) then become\footnote{(25) and (26) imply that Feynman's \cite{Fey} cascade scenario whereby a vortex ring breaks up into smaller and smaller rings is apparently inconsistent with the simultaneous conservation of energy and momentum (Svistunov \cite{Svi}) because a decay into arbitrary small rings with the total vortex line length conserved (to conserve total energy) would lead to vanishing total momentum.}
\be\tag{25}
E = \frac{1}{2} \int A_\theta \Omega_\theta d {\bf s} =  \pi R \kappa A_\theta (R)
\ee
\be\tag{26}
{\bf P} = P {\ihat}_z, ~P = \frac{1}{2}\int R \Omega_\theta d {\bf s} = \pi R^2 \kappa.
\ee

If the total kinetic energy of the superfluid is taken to be proportional to the vortex line density, then $E \sim R$, and (25) implies,
\be\tag{27}
A_\theta (R) \sim const
\ee
and (26) implies in turn,
\be\tag{28}
E \sim \sqrt{P}.
\ee

If $Q$ is the coordinate conjugate  to $P$, Hamilton's equation is
\be\tag{29}
\frac{d Q}{d t} = \frac{\pa E}{\pa P}
\ee
and using (28) and (26), (29) becomes
\be\tag{30}
\frac{d Q}{d t} \sim \frac{1}{\sqrt{P}} \sim \frac{1}{R}.
\ee

If we take next $Q$, (which is a measure of the distance traversed by the vortex ring along its axis) to be
\be\tag{31}
Q \sim R^{-2}
\ee
(30) becomes
\be\tag{32}
\frac{d R}{d t} \sim -R^2
\ee
which is just the vortex line decay equation (18), underscoring the conservative nature of the mechanism underlying vortex line decay at low temperatures.

(30) further implies that vortex rings, as is well known (Rayfield and Reif \cite{Ray}), propagate faster as they shrink, as per (32).\footnote{It is interesting to note that in an ordinary fluid, by contrast, the circulation around a vortex decreases (due to vorticity loss via detrainment of the vortical fluid into the wake) with a concomitant increase in the radius of the ring, as per (26), and a slowing down of the propagation of the ring, as confirmed by the laboratory experiments (Maxworthy \cite{Max}).}

\vspace{.10in}

\noindent\Large\textbf{5. Finite-time Singularity in the Velocity Field}\\

\large The complexity of the vortex line field (Constantin \cite{Con}) is believed to be connected with the {\it finite-time} singularity (FTS)\footnote{It may be mentioned, however, that, for ordinary fluids, there is no conclusive numerical evidence (Brachet et al. \cite{Bra3}) that ideal-flow solutions, starting from regular initial conditions, will spontaneously develop a singularity in {\it finite} time.} development, if any, in classical turbulence. The mechanism of dynamic tangle of vortex lines in superfluid turbulence is believed to generate (as the mechanism of vortex stretching does in classical turbulence) strongly localized features in the small-scale structure as well as singularities in the velocity field.

In order to see this, note that (8a), (9) and (13) give
\be\tag{33}
\eps \sim \kappa \Omega^2 \sim const.
\ee

On the other hand, using (16), one may write for the vorticity evolution,
\be\tag{34}
\frac{d \Omega}{d t} \sim \kappa \frac{\Omega}{\ell^2}
\ee
which, on using (10), becomes
\be\tag{35}
\frac{d \Omega}{d t} \sim \sqrt{\frac{\eps}{\kappa}} \Omega.
\ee
(35), in turn, on using (33), becomes
\be\tag{36}
\frac{d \Omega}{d t} \sim \Omega^2
\ee
as in classical turbulence (Leray \cite{Ler}) (which is plausible because of the prevalence of Kolmogorov spectrum (12) in superfluid turbulence). (36) leads to
\be\tag{37}
\Omega (t) \sim \frac{1}{t + c}
\ee
exhibiting a FTS; $c$ is an arbitrary constant.\footnote{Alternatively, (37) also follows by using (21) in (13).}

\vspace{.3in}
\pagebreak

\noindent\Large\textbf{6. Spatial Intermittency Effects}\\

\large The inertial range formulations discussed in Sections 2-5 do not take into account the spatial intermittency in superfluid turbulence that was revealed by the laboratory experiments (Salort et al. \cite{Sal3}, Paoletti et al. \cite{Pao}).\footnote{Spatial intermittency in superfluid turbulence was observed in the quasi-classical regime as well (Maurer and Tabeling \cite{Mau}).} The underlying cause for spatial intermittency appears to be excessive vortex line crinkling operational at length scales smaller than $\ell$, as a consequence of which, the vortex lines are not smooth in this range (Tsubota et al. \cite{Tsu2}, Vinen \cite{Vin2}). One may therefore follow Mandelbrot \cite{Man} and argue that the spatial intermittency effects in superfluid turbulence are related to the fractal nature of the vortex lines,\footnote{The numerical simulations (Sasa et al. \cite{Sas}) of quantum turbulence  in a Bose-Einstein condensate modeled by the Gross-Pitaevskii equation also confirmed self-similar structures of tangled vortex filaments.} for length scales of the order or smaller than $\ell$.

Suppose $D \lp 1 \leq D \leq 3 \rp$ is the fractal dimension \footnote{$D$ relates the length of a vortex line with its three-dimensional extent.} of a vortex line in superfluid turbulence. Then, the intervortical space filling factor $\beta$ is given by
\be\tag{38}
\beta \sim \ell^{2 - f (D)}
\ee
where $f (D)$ may be interpreted as being the fractal dimension of the support of the measure in question, and satisfies the following properties,
\be\tag{39}
\begin{aligned}
& * f (D) \geq 0, ~1 \leq D \leq 3\\
& * f^\pr (D) < 0, ~1 \leq D \leq 3\\
& * f (1) = 2,\\
& * f (3) = 0.
\end{aligned}
\ee
(39) implies,
\be\tag{40}
f (D) = 3 - D.
\ee

Using (40), (38) gives
\be\tag{41}
\beta \sim \ell^{\lp D - 1 \rp}.
\ee

On the other hand, the total length $\mathcal{L}$ of the vortex line is given by
\be\tag{42}
\mathcal{L} \sim L \beta V \sim \frac{V}{\ell^2}
\ee
$V$ being the volume of the region occupied by the superfluid. Using (41), (42) gives
\be\tag{43}
L \sim \ell^{-\lp D + 1 \rp} ~\text{or} ~\ell \sim L^{-\lp \frac{1}{D + 1} \rp}
\ee
which reduces to (1) in the smooth vortex-line limit $D \Rightarrow 1$.

(43) may be rewritten as
\be\tag{44a}
L \sim \ell^{- \lp D - 1 \rp} \cdot \ell^{-2}
\ee
or in Vinen's \cite{Vin2} notation,
\be\tag{44b}
L \sim g \cdot \ell^{-2}
\ee
where the {\it intermittency correction factor} $g$ is given by
\be\tag{45}
g \sim \ell^{- \lp D - 1 \rp} > 1.
\ee

(44) implies, as Vinen \cite{Vin2} predicted, the enhancement of the vortex line density $L$, caused by the excessive crinkling of the vortex lines in the Kelvin wave cascade. The vortex line density enhancement in superfluid turbulence was confirmed by laboratory experiments (Walmsley et al. \cite{Wal}) and is commensurate with the enhanced depolarization of vortex lines (as also confirmed again in Section 6 (ii)) in a more dense vortex tangle.

\vspace{.3in}

\noindent\textbf{(i) Energy Spectrum}\\

The energy per unit mass at length scale $\ell$ in the presence of spatial intermittency (following the $\beta$-model of Frisch et al. \cite{Fri} in classical turbulence) is
\be\tag{46}
E (\ell) \sim \beta\frac{\kappa^2}{\ell^2}
\ee
so the energy transfer rate per unit mass at length scale $\ell$ is (on using (7)),
\be\tag{47}
\eps (\ell) \sim \frac{E (\ell)}{t (\ell)} \sim \beta \frac{\kappa^3}{\ell^4}.
\ee
On using (41), (47) becomes
\be\tag{48}
\eps (\ell) \sim \kappa^3 \ell^{\lp D - 5 \rp}.
\ee

Constancy of the energy transfer rate in the cascade, namely (9), then gives
\be\tag{49a}
\kappa \sim \eps^{1/3} \ell^{\lp \frac{5-D}{3} \rp}
\ee
or
\be\tag{49b}
\ell \sim \frac{\kappa^{\lp \frac{3}{5-D} \rp}}{\eps^{\lp \frac{1}{5-D} \rp}}
\ee
which may be rewritten as
\be\tag{49c}
\ell \sim \frac{\kappa^{\frac{3}{4} \lb 1 + \lp \frac{D - 1}{5-D} \rp \rb}}{\eps^{\frac{1}{4} \lb 1 + \lp \frac{D - 1}{5-D} \rp \rb}}.
\ee
(49c) reduces to (10) in the smooth vortex-line limit $D \Rightarrow 1$.

Using (41) and (49), (46) gives
\be\tag{50}
E (\ell) \sim \eps^{2/3} \ell^{1/3 \lp D + 1 \rp}
\ee
which leads to the energy spectrum,
\be\tag{51}
E (k) \sim \eps^{2/3} k^{-5/3 - 1/3 \lp D - 1 \rp}.
\ee

Observe that, since $D > 1$, (51) implies that the spatial intermittency effects make the energy spectrum steeper (as is also the case in classical turbulence), in qualitative agreement with the laboratory experiment results (Salort et al. \cite{Sal3}) on spatially intermittent superfluid turbulence.\footnote{Similar qualitative results were also observed in the quasi-classical regime (Maurer and Tabeling \cite{Mau}).}

\vspace{.3in}

\noindent\textbf{(ii) Vortex Line Decay}\\

Using (41) and (43), (46) and (47) become 
\be\tag{52a}
E (L) \sim \kappa^2 L^{-\lp \frac{D - 3}{D + 1} \rp}
\ee
\be\tag{53a}
\eps (L) \sim \kappa^3 L^{-\lp \frac{D - 5}{D + 1} \rp}.
\ee

(53a) may be rewritten (in the notation of Vinen \cite{Vin8}) as
\be\tag{53b}
\eps (L) \sim \nu^{\pr \pr} (L) \kappa^2 L^2
\ee
where,
\be\tag{53c}
\nu^{\pr \pr} (L) \sim \displaystyle\kappa L^{-3 \lp \frac{D - 1}{D + 1} \rp}
\ee
confirming the weak dependence on the parameter $\nu^{\pr \pr}$ on $L$ conjectured by Vinen \cite{Vin8}.

Substituting (52a) and (53a) into (17), we obtain for the vortex line decay,
\be\tag{54a}
\frac{d L}{d t} \sim -\kappa L^{\lp \frac{D + 3}{D + 1} \rp}
\ee
or
\be\tag{54b}
\frac{d L}{d t} \sim -\kappa L^{2 -  \lp \frac{D - 1}{D + 1} \rp}.
\ee
(54) gives,
\be\tag{55a}
L (t) \sim t^{-\lp \frac{D + 1}{2} \rp}
\ee
or
\be\tag{55b}
L (t) \sim t^{-1 - 1/2 \lp D - 1 \rp}.
\ee
(55) (or (54)) shows enhancement of the vortex line decay due to spatial intermittency effects $\lp D > 1 \rp$ and appears to be consistent with the belief that an increase in polarization of vortex lines in a less dense vortex tangle inhibits vortex line reconnections (Laurie et al. \cite{Lau}) and leads to a reduction in the rate of vortex-line decay (Vinen \cite{Vin4}).\footnote{It is of interest to note that, corresponding to $D = 5/3$, ($D = 5/3$ corresponds to the value of fractal dimension of a vortex tangle as if it is a self-avoiding walk (de Gennes \cite{deG}) (55) yields $L\lp t\rp\sim t^{-4/3}$, while $L\lp t\rp\sim t^{-3/2}$ was observed in the laboratory experiments (Walmsley et al. \cite{Wal}) on the decay of a vortex tangle in the He-II in a closed tube.}

(55a) further implies that the intervortical spacing scales according to
\be\tag{56}
\ell (t) \sim L^{-\lp\frac{1}{D+1}\rp}\sim t^{\frac{1}{2}}
\ee
\noindent implying that spatial intermittency apparently has no effect on the intervortical spacing enhancement rate. Some insight into this may be gained by noting that spatial intermittency leads to two mutually opposing effects,
\begin{itemize}
\item kinematical (due to enhanced vortex line crinkling), which decreases the intervortical spacing;
\item dynamical (due to vortex line decay) which increases the intervortical spacing;
\end{itemize}
which apparently cancel each other.

On the other hand, (52a) shows that $E (L)$ is no longer proportional to $L$ in the presence of spatial intermittency $\lp D > 1 \rp$. However, one may define a renormalized vortex line density $\bar{L}$ to incorporate spatial intermittency effects according to\footnote{A similar smoothing of $L$ to incorporate the effects of fractality of vortex filaments was proposed by Kozik and Svistunov \cite{KozSvi}, while Vinen \cite{Vin4} had proposed this idea to incorporate the effects of Kelvin waves..}
\be\tag{57a}
\bar{L} \equiv L^{\lp \frac{3 - D}{D + 1} \rp}
\ee
or
\be\tag{57b}
\bar{L} \equiv L^{1 - 2 \lp \frac{D - 1}{D + 1} \rp} < L
\ee
which also seems to be plausible due to reduced polarization of vortex lines in the spatially intermittent case. (52a) then yields,
\be\tag{52b}
E \lp \bar{L} \rp \sim \kappa^2\bar{L}
\ee
which has the same form as that for the non-intermittent case.

We have from (55) and (57) that
\be\tag{58a}
\bar{L} \sim t^{-1/2 \lp 3 - D \rp}
\ee
or
\be\tag{58b}
\bar{L} \sim t^{-1 + 1/2 \lp D-1 \rp}
\ee
which implies that $\bar{L}$ decays slower in the spatially intermittent case $\lp D > 1 \rp$ due to an effectively increased polarization of vortex lines associated with $\bar{L}$.

\vspace{.3in}

\noindent\textbf{(iii) Finite-time Singularity in the Velocity Field}\\

The spatial intermittency effects on the FTS development, as in classical turbulence (Shivamoggi \cite{Shi4}), may be expected to materialize in superfluid turbulence as well.

We have from (47), (13) and (9),
\be\tag{59}
\eps \sim \beta \kappa \Omega^2 \sim \beta \frac{\kappa^3}{\ell^4} \sim const.
\ee

Assuming the scaling behavior,
\be\tag{60}
\kappa \sim \ell^\al
\ee
and using (41), (59) gives
\be\tag{61}
3 \al +  \lp D - 1 \rp - 4 = 0
\ee
from which,
\be\tag{62}
\al = \frac{5 - D}{3}
\ee
in agreement with (49a).

Using (60) and (62), the vorticity evolution equation (34) gives
\be\tag{63}
\frac{d \Omega}{d t} \sim \kappa^{-\lp \frac{D + 1}{5-D} \rp} \Omega
\ee
and using (33), (63) becomes
\be\tag{64}
\frac{d \Omega}{d t} \sim\varepsilon^{-\lp\frac{D+1}{5-D}\rp} \Omega^{\lp \frac{7+D}{5 - D} \rp}
\ee
from which,
\be\tag{65a}
\Omega (t) \sim \lp t + c \rp^{-1/2 \lp \frac{5-D}{D + 1} \rp}
\ee
or
\be\tag{65b}
\Omega (t) \sim \lp t + c \rp^{-1 + 3/2 \lp \frac{D - 1}{D + 1} \rp}.
\ee
(65) shows a weakening of the finite-time singularity by the spatial intermittency effects, which may be traced to the enhanced vortex line decay (found above in Section 6 (ii)) due to the latter. This is similar to the situation in classical turbulence (Shivamoggi \cite{Shi4}) where nonlinearity depletion mechanisms via coherent structure generation have been speculated to be operational (Frisch \cite{Fri2}).

\vspace{.3in}

\noindent\Large\textbf{7. Discussion}\\

\large One of the major issues in theoretical investigations on the superfluid turbulence problem has to do with the expectation that superfluid turbulence, at low temperatures $(T < 1K)$ and length scales less than the intervortical distance $\ell$, would be very different from classical turbulence (at these low temperatures, the normal fluid component essentially vanishes while the Kelvin waves govern the dynamics at these small length scales) whereas laboratory experiments (Salort et al. \cite{Sal3}) and numerical simulations (Araki et al. \cite{Ara}) indicated otherwise\footnote{It is of interest to note that similar situations arise elsewhere. The energy spectrum in electron magnetohydrodynamics (EMHD) turbulence turns out (Biskamp et al. \cite{Bis}, Shivamoggi \cite{Shi5}) to follow the Kolmogorov spectrum, in the large wavenumber limit, in spite of the fact that the whistler waves (which are generic to EMHD) control the underlying cascade physics.}. In recognition of this, theoretical considerations are made in this paper of superfluid turbulence in the Kelvin wave cascade regime at low temperatures and small length scales that confirm an energy spectrum exhibiting the Kolmogorov scaling observed in laboratory experiments (Salort et al. \cite{Sal3}) and numerical simulations (Araki et al. \cite{Ara}). The vortex line decay mechanism at low temperatures is conservative in nature, and can hence be characterized via a Hamiltonian framework. 

Further, laboratory experiments (Maurer and Tabeling \cite{Mau}, Salort et al. \cite{Sal3}) gave evidence of inertial range spatial intermittency in superfluid turbulence. On the other hand, because of excessive crinkling occuring at small length scales, the vortex lines become non-smooth and fractal-like in this range (Tsubota et al. \cite{Tsu2}, Vinen \cite{Vin2}). In recognition of this, in this paper, spatial intermittency effects are then incorporated into the present theoretical formulations, following Mandelbrot \cite{Man}, via the fractal nature of the vortex lines. The latter aspect is shown to enhance the vortex line density $L$, for a given value of intervortex spacing $\ell$ (as conjectured by Vinen \cite{Vin2}) and to provide for a mechanism commensurate with the enhanced depolarization of vortex lines. The spatial intermittency is found to steepen the energy spectrum in qualitative agreement with the laboratory experiments (Salort et al. \cite{Sal3}) and to enhance vortex line decay in agreement with the remarks of Vinen \cite{Vin4}.

It remains to be emphasized, however, that there are still many subtle characteristics of the superfluid physics that have been evaded in the present theoretical discussions.

\vspace{.3in}

\noindent\Large\textbf{Acknowledgments}\\

\large This work was carried out during my visiting appointment at the International Centre for Theoretical Sciences, Bengaluru. I am thankful to Professor Spenta Wadia for his hospitality. I am thankful to Professor Katepalli Sreenivasan for his constant encouragement, helpful remarks and suggestions. I am thankful to Professor Predhiman Kaw for helpful discussions. I am thankful to Dr. Demosthenes Kivotides for his helpful suggestions and criticism.

\end{document}